\documentclass[useAMS,usenatbib]{mn2e}
\usepackage{graphicx,amsmath}

\title[The full distribution function in halos]{Analytical derivation of the radial distribution function in spherical dark matter halos}
\author[Eilersen et al]{Andreas Eilersen, Steen H. Hansen, Xingyu Zhang\\
Dark Cosmology Centre, Niels Bohr Institute, University of Copenhagen,
  Juliane Maries Vej 30, 2100 Copenhagen, Denmark}
\begin{document}
\pagerange{\pageref{firstpage}--\pageref{lastpage}} \pubyear{2012}

\maketitle

\label{firstpage}

\begin{abstract}
The velocity distribution of dark matter near the Earth is important
for an accurate analysis of the signals in terrestrial detectors. This
distribution is typically extracted from numerical simulations. Here
we address the possibility of deriving the velocity distribution
function analytically. We derive a differential equation which is a
function of radius and the radial component of the velocity. Under
various assumptions this can be solved, and we compare the solution
with the results from controlled numerical simulations. Our findings
complement the previously derived tangential velocity distribution.
We hereby
demonstrate that the entire distribution function, below $\sim 0.7
v_{\rm esc}$, can be derived analytically for spherical and
equilibrated dark matter structures.
\end{abstract}


\section{Introduction}
The many gravitational observations of dark
matter~\citep{2006ApJ...648L.109C, 2015arXiv150201589P} makes us eager
to measure the dark matter either in underground observatories or
indirectly from the decay or annihilation of dark matter. The analysis
of a potential signal in underground observatories requires an
assumption of the velocity distribution function of the dark matter
near the Earth \citep{2013EPJC...73.2648B,2016PhRvL.116p1301A},
however, this velocity distribution remains unclear. This problem has
in particular been addressed via numerical simulations (for a list of
recent references, see
\cite{2015MNRAS.452..944B,2016arXiv160505185C,2016arXiv160104725K,2014ApJ...784..161P}).

A wide range of theoretical ideas have been suggested to understand
the dynamical origin of the distribution function
e.g. \citep{2003A&A...408...27L, 2004ApJ...604...18W,
  2006MNRAS.368.1931L, 2007ApJ...666..181S, 2007MNRAS.376..393A,
  2010arXiv1010.2539D}.  Also several papers have attempted
statistical mechanical approaches, e.g. \cite{1957SvA.....1..787O,
  1967MNRAS.136..101L, 2005NewA...10..379H, 2010ApJ...722..851H}.

The most dominating guiding principle for theoretical investigations
has been the Jeans theorem, which states that {\em any} function of
the integrals of motion yields a solution to the steady-state
Boltzmann equation~\citep{binneytremaine}. In other words, any
function of energy and angular momentum, $f(E,L)$, will be a solution
to the collisionless Boltzmann equation. This has lead to a large
number of authors studying exactly such
functions~\citep{1916MNRAS..76..572E,2005MNRAS.360..492E,2006PhRvD..73b3524E,2006ApJ...642..752A,2014ApJ...783...13W,2008MNRAS.388..815W}.
An alternative approach is to separate the cosmological structure in
radial bins, and then consider the distribution of velocities in each
bin. This way it was shown that the shape of the distribution function
found in numerically simulated strutures is highly non-trivial and
differs from a Gaussian~\citep{2006JCAP...01..014H}. In \cite{2015MNRAS.452..944B}
it was found that a non-separable distribution function was linearly
related to the slope of the density profile.

In this paper we will take a new approach, in order to analytically
derive the distribution function. We will attempt to derive the shape
of the radial components of the distribution function.  The idea is to
integrate the Boltzmann equation over all the variables we are not
interested in, leaving us with a differential equation,
Eq.~(\ref{eq:radvel}), over the variables $r$ and $v_r$. We will solve
this equation under various assumptions, and we can finally compare
the solution with the results of numerically simulated halos of
collisionless particles.

Our conclusions are, that when we are given the density profile of a
dark matter halo, then we can analytically derive both the radial and
tangential distribution function, and hence we know the full
distribution function of dark matter particles. The comparisons with
numerically simulated structures show that this holds true for velocities
below $\sim 0.7$ times the escape velocity.

\section{Deriving the radial equation}

The Boltzmann equation is the differential
equation describing the flow of particles in phase space~\citep{binneytremaine}:
\begin{equation}
\frac{df}{dt} =
 \frac{\partial f}{\partial t}+ \vec{v} \cdot \nabla f - \nabla \Phi \cdot \frac{\partial f}{\partial \vec{v}}=0 \, ,
\end{equation}
where the function $f(\vec{r},\vec{v},t)$ is the distribution function which
gives the density of particles in phase space.  In this paper we
will discuss the dependence of this function on the radial velocity
component $v_r$.

Solving this equation is impossible in most cases, and one therefore
has to make approximations. The most well-known approach is to get
rid of virtually all the velocity information, by integrating the
Boltzmann equation over all velocities, for instance
\begin{equation}
\int v_r \, \frac{df}{dt} \, d^3v = 0 \, ,
\end{equation}
which leads to one of the Jeans equations
\begin{equation}
\frac{GM(r)}{r}=-\overline{v_r^2} \left( \frac{d {\rm ln}(\rho)}{d
  {\rm ln} (r)}+\frac{d {\rm ln}(\overline{v_r^2})}{d {\rm ln}(r)}+2\beta \right) \, ,
\label{eq:jeans}
\end{equation}
where $\beta \equiv 1-\frac{\overline{v_\theta^2}}{\overline{v_r^2}}$
and we have assumed $\overline{v_\phi^2}=\overline{v_\theta^2}$. We
have also assumed sphericity, such that
$\frac{d\Phi}{dr}=\frac{GM(r)}{r^2}$.  This is the Jeans equation most
commonly used for describing collisionless structures like galaxy
clusters and dwarf galaxies.

\subsection{Deriving an equation for the radial VDF}

The problem with the Jeans
equation is that, by averaging over the entire velocity
space in calculating the radial velocity dispersion $\sigma_r^2 \equiv
\overline{v_r^2}$, we lose information about the detailed
shape of the velocity distribution function (VDF).

We will therefore integrate the Boltzmann
equation only over the two angular velocity components instead of all
three $v_i$'s
\begin{equation}
\int \frac{df}{dt} \, d^2v_{\theta,\phi} = 0 \, .
\end{equation}
Thereby the radial velocity will remain a free
variable. The result, as we shall see, is a differential equation
in the variables $v_r$ and $r$

We again start with the Boltzmann equation
under the assumption of spherical symmetry and staticity:
\begin{equation}
v_r \frac{\partial f}{\partial r}+\left(\frac{v_\theta^2+v_\phi^2}{r}-\frac{\partial \Phi}{\partial r}\right)\frac{\partial f}{\partial v_r}-\frac{v_rv_\theta}{r}\frac{\partial f}{\partial v_\theta}-\frac{ v_r v_\phi}{r}\frac{\partial f}{\partial v_\phi}=0
\end{equation}

Integrating over the two angular velocity components in phase space
gives us

\begin{eqnarray}
v_r \frac{\partial}{\partial r} \int f dv_{\theta , \phi}+\frac{1}{r} \frac{\partial }{\partial v_r} \int (v_\theta^2+v_\phi^2)fdv_{\theta , \phi} && \nonumber \\
-\frac{\partial \Phi}{\partial r} \frac{\partial }{\partial v_r}  \int fdv_{\theta , \phi}+2\frac{v_r}{r} \int f dv_{\theta ,\phi}&=&0
\end{eqnarray}

To simplify this equation we define $F_R \equiv \int f dv_{\theta ,
  \phi}$, which is the density of particles in a given volume of space
with a given radial velocity $v_r$.  We also define $\langle
v_i^2\rangle_r \equiv \int fv_i^2dv_{\theta , \phi} $, which is the
weighted sum of squared i-velocity components in the phase plane
($v_\theta$ , $v_\phi$), with f acting as weight. If divided by $F_R$,
this would be completely analogous to the averages $\overline{v_i^2}$
in the Jeans equation. Using this, our equation becomes
\begin{equation} 
v_r \frac{\partial F_R }{\partial r} + \frac{1}{r} \frac{\partial
}{\partial v_r} \left( \langle v_\theta^2\rangle _r+\langle
v_\phi^2\rangle _r \right) -\frac{d \Phi }{d r} \frac{\partial F_R
}{\partial v_r}+2\frac{v_r \hspace{1mm} F_R}{r}=0 \, .
\label{eq:radvel}
\end{equation}

This is a differential equation in two variables $v_r$ and $r$, and
with four unknown functions, $F_R$, $\langle v_\theta^2\rangle _r$, $\langle v_\phi^2\rangle _r$
and $\Phi (r)$. As it is, the equation has far too
many unknowns to be solved directly. 

In order to proceed to solve this differential equation, we will assume that
the distribution function can be separated into the components
\begin{equation}
f(r,v_r,v_t) \propto f_t(v_t,k(r)) \times R(r) \times f(v_r) \, ,
\end{equation}
where $k(r)$ is an $r$-dependent normalisation parameter.  Next we
need to calculate approximate expressions for $F_R$ and $\langle
v_i^2\rangle _r$ as functionals of $R(r)f(v_r)$. This will be done
using the shape of the tangential velocity distribution derived in
\cite{2005NewA...10..379H}, which was demonstrated to be in good
agreement with numerical simulations  \citep{2012ApJ...756..100H}.
In this way, the number of unknown functions in the above equation can
be reduced to two, namely $f_r(v_r)$ and $R(r)$, and the equation can
then be solved using regular separation of variables.

\subsection{The tangential VDF}
Historically, attempting to derive
a general expression for VDFs in static, spherical systems, Eddington realised
that in order for the system to be static, the inflow
from an outer spherical shell 
must equal the flow out of the smaller inner radial bin. The flux of
particles as a function of average velocity clearly also depends on
the density in the various bins, and therefore the density profile of
the entire structure. Using the requirement of staticity, Eddington
was able to relate the velocity distribution function to the density
profile of the spherically symmetric structure \citep{1916MNRAS..76..572E}
\begin{equation}
f(E)=\frac{1}{\sqrt{8}\pi}\int_0^E \frac{d^2 \rho}{d \Psi^2} \frac{d\Psi}{\sqrt{E-\Psi}} \, ,
\end{equation}
where $-\Psi (r)=\Phi (r)$ is the potential, $\rho (r)$ is
the density profile, and $E=\Psi -\frac{v^2}{2}$ is the relative
energy. We have also assumed $\beta =0$. This method for finding the
velocity distribution function is know as the \emph{Eddington
  inversion method}~\citep{binneytremaine}.
One unfortunate problem with this method is, that the resulting distribution
functions do not agree with the ones observed in numerical simulations.

It has been suggested that the tangential component of the
distribution function should be particularly
simple~\citep{2005NewA...10..379H}.  For sufficiently small
velocities, the tangential velocity components only cause the
particles to move around within the same radial bins. Given spherical
symmetry, the potential and density at constant $r$ are themselves
constant. In this vastly simplified case, the Eddington inversion
method yields
\begin{equation}
f(v_t) \propto \left( 1+\frac{v_\theta^2+v_\phi^2}{3k^2}\right)^{-5/2} \, .
\label{eq:ftan}
\end{equation}

This function is known as a Tsallis distribution and has wide
applications in non-extensive statistical mechanics, where functions
of this type tend to maximise the generalised entropy. Clearly, the
resulting
analytical distribution function has a long tail of high
energy particles, which cannot exist in finite structures, where the
maximum velocity at any radius is the escape velocity, $v_{\rm esc} =
\sqrt{2\left| \Phi \right|}$. This immediately tell us, that this
function cannot provide a good fit at high velocities. Surprisingly
enough, this function gives an excellent fit for all velocities
smaller than $0.7 \, v_{\rm esc}$ at all radii. This holds for
structures formed in controlled numerical simulations (spherical
collapse, head-on collisions, and various perturbed structures) and
also in cosmological simulated structures inside a density slope
around $\gamma = -2.5$.

\section{Solving the radial equation}

As clarified above, we need to make two assumptions to solve
the equation for the radial distribution function, Eq.~(\ref{eq:radvel}), namely that the
full distribution function approximately is separable, 
$f(r, v_t,v_r)=R(r) \times f_t(v_t) \times f_r(v_r)$, and that the shape of the tangential distribution
funtion is known. Substituting the tangential
VDF in Eq.~(\ref{eq:ftan}), we get
\begin{equation}
f(\vec{v},r) \propto \left(1+\frac{v_\theta^2+v_\phi^2}{3k^2}\right)^{-5/2} R(r)f (v_r) \, .
\end{equation}
For ease of notation we denote the
$v_r$-component of the function as $f(v_r)$ as opposed to the full
function $f(v_r,r)$. We then calculate $\langle v_i^2\rangle _r$ and $F_R$ 
\begin{equation}
F_R \propto \int_{-\infty}^{\infty} \int_{-\infty}^\infty \left(1+\frac{v_\theta^2+v_\phi^2}{3k^2}\right)^{-5/2}R(r) f (v_r)dv_\theta dv_ \phi \, ,
\end{equation}
\begin{equation}
\langle v_\theta^2\rangle _r \hspace{1mm} \propto  \int_{-\infty}^{\infty} \int_{-\infty}^\infty \left(1+\frac{v_\theta^2+v_\phi^2}{3k^2}\right)^{-5/2} v_\theta^2 f (v_r)R(r)dv_\theta dv_ \phi \, .
\end{equation}

These integrals give us
\begin{equation} \label{eq:Fr-relation}
F_R \propto 2 \pi k^2(r)f(v_r)R(r) \, ,
\end{equation}

\begin{equation} \label{eq:vtan-relation}
\langle v_\theta^2\rangle _r =\langle   v_\phi^2\rangle _r \hspace{1mm} \propto 6 \pi k^4(r) f(v_r)R(r) \, .
\end{equation}

Plugging these into the radial velocity equation, we get
\begin{equation}
\frac{R'(r)}{R(r)}+\frac{1}{k^2(r)}\frac{dk^2(r)}{dr}+\frac{2}{r}
=\left(\frac{GM(r)}{r^2}-\frac{6k^2(r)}{r}\right) \frac{f'(v_r)}{v_rf(v_r)} \, .
\end{equation}

Separation implies that we have the two ordinary differential equations
\begin{equation}
\frac{\frac{R'(r)}{R(r)}+\frac{1}{k^2(r)}\frac{dk^2(r)}{dr}+\frac{2}{r}}{\frac{GM(r)}{r^2}-\frac{6k^2(r)}{r}}=b_c \, ,
\end{equation}
\begin{equation}
\frac{f'(v_r)}{v_rf(v_r)}=b_c \, ,
\end{equation}
with the solutions
\begin{equation}
R(r)=c_2 {\rm exp} \left[ - \int \left(b_c\left(6k^2(r) - v_c^2\right) + 2 + \frac{d {\rm ln} k^2(r)}{d {\rm ln} r}\right) d{\rm ln}r \right] \, ,
\end{equation}
\begin{equation}
f(v_r)=c_3 e^{\frac{1}{2}b_cv_r^2} \, ,
\label{eq:frad}
\end{equation}
where we have that $f(v_r,r)\propto R(r)f(v_r)$. 

\begin{figure}
        \includegraphics[angle=0,width=0.49\textwidth]{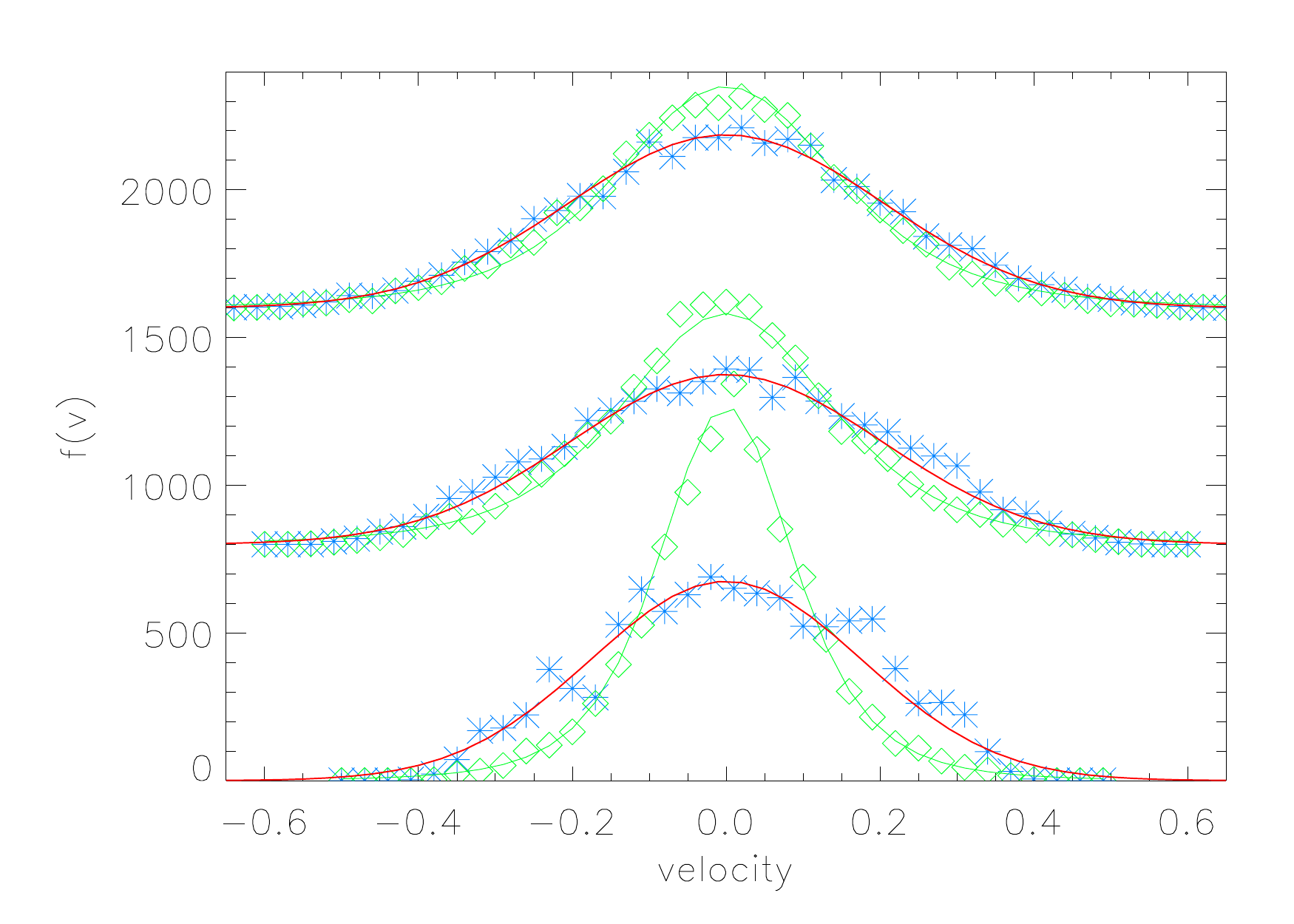}
\caption{The radial (blue stars) and tangential (green diamonds)
  velocity distribution at 3 different radii (corresponding to $\gamma
  = d {\rm ln} \rho/d {\rm ln} r = -1.8, -2.3, -3$) after repeated
  perturbations of the gravitational potential. The figures are
  shifted vertically to improve readability. The radial distributions
  are all fitted with the shape in Eq.~(\ref{eq:frad}) (thick red
  lines), and the tangential distributions are all fitted with the
  shape in Eq.~(\ref{eq:ftan}) (thin green lines). When plotted in
  lin-lin the fits appear rather good.}
\label{fig:fig1}
\end{figure}

\begin{figure}
        \includegraphics[angle=0,width=0.49\textwidth]{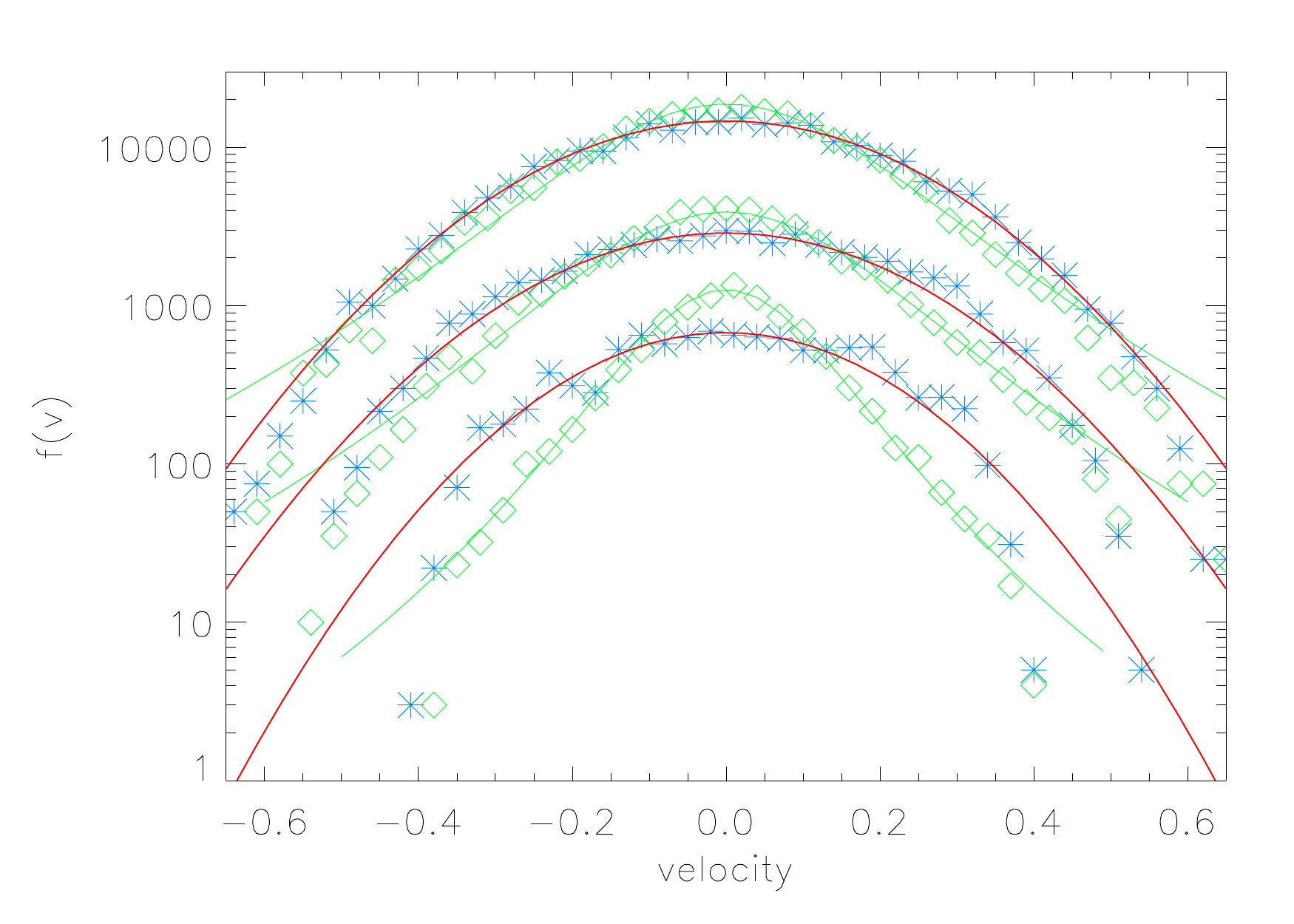}
\caption{The radial (blue stars) and tangential (green diamonds)
  velocity distribution at 3 different radii (corresponding to $\gamma
  = -1.8, -2.3, -3$) after repeated perturbations of the gravitational
  potential. The figures are shifted vertically to improve
  readability. The radial distributions are all fitted with the shape
  in Eq.~(\ref{eq:frad}) (thick red lines), and the tangential
  distributions are all fitted with the shape in Eq.~(\ref{eq:ftan})
  (thin green lines). When plotted in lin-log it is clear that the
  low-velocity region is fairly well fitted, whereas the high-velocity
  region is not.}
\label{fig:fig2}
\end{figure}

\section{Comparison with numerical simulations}

In order to test the accuracy of our solution, we compare with the
velocity distributions extracted from a range of controlled numerical
simulations. The reasons for using controlled (instead of cosmological)
simulations is that we thereby can assure that equilibrium is 
achieved. When a structure has reached equilibrium we divide it in
radial bins, and then we plot the resulting radial and tangential
velocity distribution in each radial bin. 

{We consider 3 different perturbation schemes. The first follows a
  scheme to resemble the effect of mergers by repeatedly changing the
  energies of individual particles by changing the depth of the
  potential. Practically we vary the gravitational potential, where
  the value of G is changed by a factor of 1.1 and 0.9 from its normal
  value, each for a few dynamical times. Thereby the particles are
  accelerated/deccelerated while the system is allowed to relax
  through phase-mixing \citep{2012JCAP...10..049S}.  

Only the trustworthy region is considered, excluding the central
resolution-limited region with radii smaller than 5 times the
softening length. The outer non-fully-equilibrated region is
identified by considering regions where $\gamma = d {\rm ln} \rho/d
{\rm ln} r$ starts varying between individual perturbations.
Effectively this removes regions outside radii where the density slope
is -3.  The issue of trustworthy regions was analysed in detail in
\cite{2012JCAP...10..049S}.

In Figure 1 we show the distribution of velocities in 3 different
radial bins, where the slope of density is about $\gamma = d {\rm ln}
\rho/d {\rm ln} r = -1.8, -2.3, -3$ The bins are shifted vertically to
improve readability.  The blue stars are the radial velocities, and
the red solid line is the Gaussian in Eq.~(20). In comparison we also
plot the tangential velocity distribution (green diamonds) which is
well fitted with the shape in Eq.~(10).

Figure~2 shows the same distribution in lin-log space in order
to make the tails in the distribution much more visible.

The second perturbation scheme is a standard cold collapse: a
spherical distribution of particles are placed at rest, and the
violent relaxation during the collapse leads to a new equilibrated
structure. The resulting velocity distributions at 3 different radii
are shown in Figure 3.  The cold collapse is a single major
perturbation (as opposed to the repeated changing of G as described
above), and it is therefore less likely that a perfectly beautiful
distribution can be obtained. Infact, in Figure~3 we see that a
Gaussian is not a perfect fit, since variations are even visible by
eye on the lin-log figure.

It is important to keep in mind that in principle there are {\em no
free} parameters in the fits. The width of the distribution function
is given by the radial dispersion, and the height of the curve is
given by the number of particles in the radial bin. However, since
the curves are not fitted perfectly (the tail is wrong), we allow
both the dispersion and normalization as free fitting parameters, and
the resulting fits are a few percent different from the true values
from the simulations.

\begin{figure}
        \includegraphics[angle=0,width=0.49\textwidth]{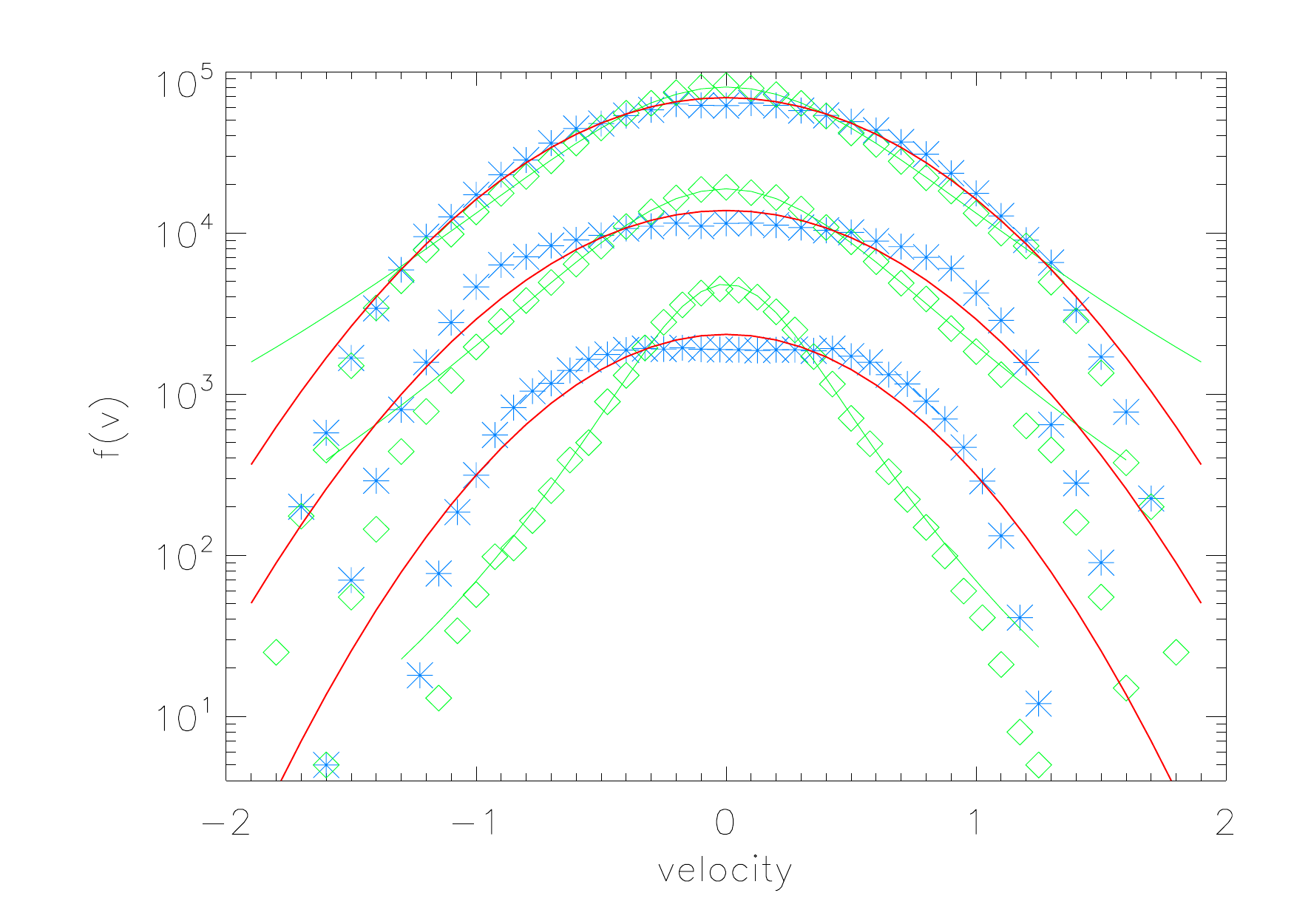}
\caption{The radial (blue stars) and tangential (green diamonds) velocity distribution at 3
  different radii (corresponding to $\gamma = -1.6, -2.0, -2.4$) after
a single spherical cold-collapse. The figures are
  shifted vertically to improve readability. The radial distributions
  are all fitted with the shape in Eq.~(\ref{eq:frad}) (thick red lines), and the
  tangential distributions are all fitted with the shape in
  Eq.~(\ref{eq:ftan}) (thin green lines). It is clear that the
low-velocity region is well fitted, whereas the high-velocity region
is not.}
\label{fig:fig3}
\end{figure}

\begin{figure}
        \includegraphics[angle=0,width=0.49\textwidth]{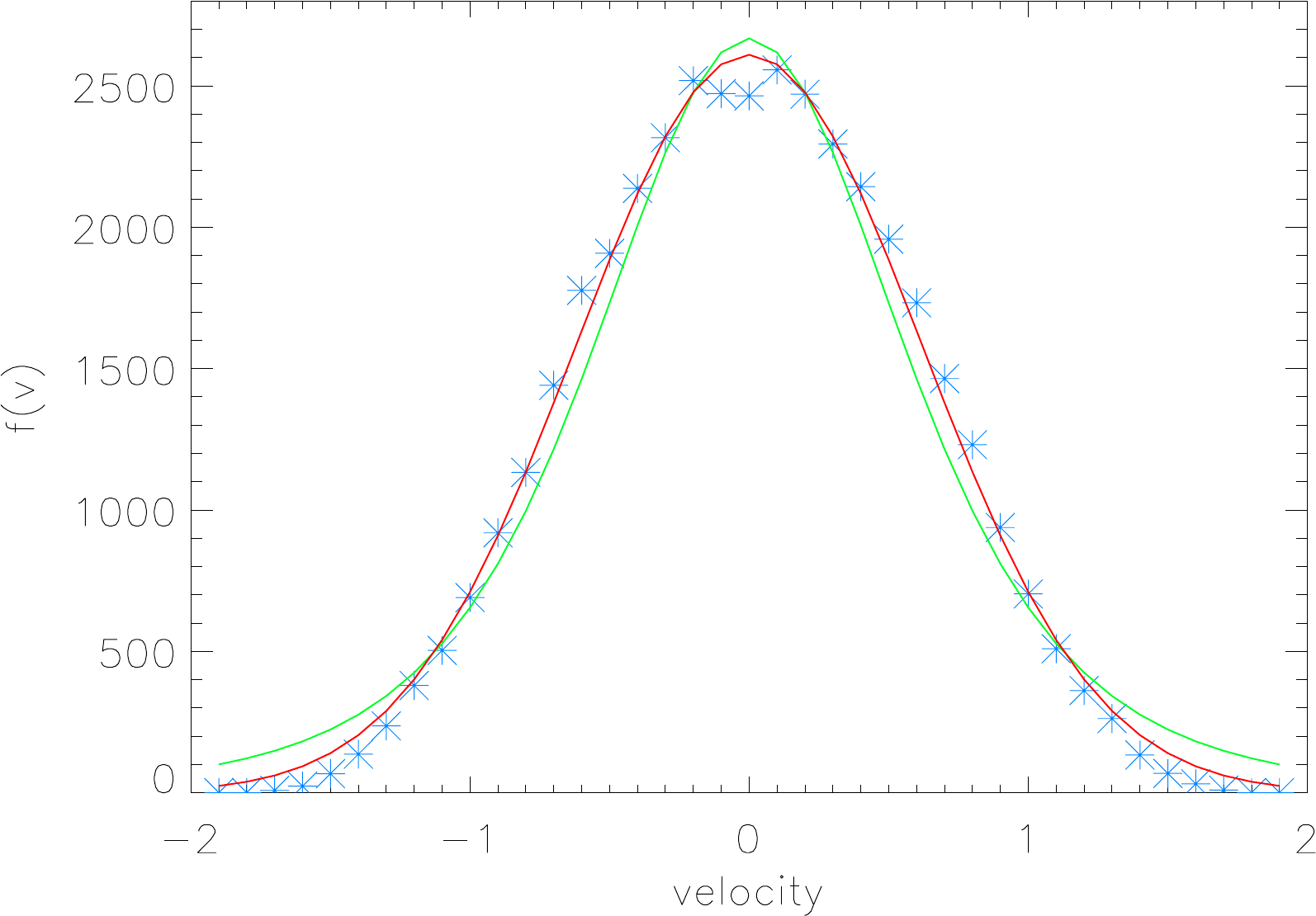}
\caption{The radial (blue stars) velocity distribution at the radius 
corresponding to $\gamma = -1.6$ after
a single spherical cold-collapse. The two lines are best fits of the Gaussian
and Tsallis shapes respectively, and it is clear that with a few percent
Poisson error-bars (which is easily achieved with 1M particles in the structure)
one can easily distinguish the two. The Gaussian is see to fit much better than
the Tsallis shape, in agreement with our analytical derivation.}
\label{fig:fig4}
\end{figure}

In order to address the question if a Gaussian is really the best fit,
we repeat the data in the lowest lines of Figure 3 in a lin-lin space,
where we show two best fit lines: the (blue) Gaussian is visibly seen
to provide a better fit than the (green) Tsallis shape: the Tsallis
shape is too low around $v \sim \sigma$, and the Tsallis is too high
at velocities above $v \approx 2\sigma$. For this radial bin the
reduced chi-squares are about 2 and 6 respectively, when fitting
velocities inside 1.8 times the dispersion (corresponding to 1.2 on
the x-axis). Another way of quantizing the preference for the Gaussian
is that the resulting error-bars on the fitting parameters (dispersion
and magnitude) are approximately a factor of two smaller when fitted
to a Gaussian than when fitted to a Tsallis shape.  The statistical
preference for the Gaussian is even bigger when fitting to higher
velocities (because the Tsallis has a much longer tail), however, that
may not be entirely reliable given the very few particles in the
high-velocity tail. With a few percent error-bars (which requires
about half a million particles in the structure) such Poisson
error-bars are easily achieved.  For the bins at largest radii there
is such a small difference between the radial and tangential
distributions, that with only half million particles in the entire
structure, the statistical error-bars would allow a fit to both a
Gaussian and a Tsallis shape, when one allows both dispersion and
magnitude as free parameters, unless one includes the highest-velocity
particles (in which case the Gaussian provides a statistically better
fit).

The last perturbation scheme is a repeated pattern of {\em kick-flow},
in the sense that first the particle energies are perturbed (to
resemble the effect of mergers) and secondly the system is allow to
relax. Each energy kick is arranged to conserve energy in each radial
bin \citep{2010ApJ...718L..68H}.  Between each kick the system is
allowed to relax through phase-mixing.

In Figure 5 we again plot the velocity distribution from 3 radial
bins, and the lower bins correspond to a region near the centre, but
well outside a region of 5 times the softening length.

}

These 3 very different perturbation schemes allow us to draw the
same conclusion: all the radial distribution function are well fitted
with the shape in Eq.~(\ref{eq:frad}) for velocities below approximately
$0.7 v_{\rm esc}$. This complements the known fact that the tangential
VDF is well fit in the same region with the form in Eq.~(\ref{eq:ftan}).
One should keep in mind, that the majority of particles are having
energies below $0.7 v_{\rm esc}$, and that the departure from the
Gaussian shape is not visible when plotting in linear-linear.

{The reason why the theoretical predictions and the numerically
  simulated structures do not agree at high velocities is unknown to
  us. In the derivation we have assumed that the tangential velocities
  follow the shape in Eq.~(10), which is known to be violated at high
  velocities. Assuming a cut-off in the tangential shape does not
  change the shape of the resulting radial distribution, but only a
  marginally different value for the radial dispersion. We therefore
  believe the origin should lie in the assumption of separability of
  the full distribution function, see Eq.~(11). The full differential
  equation~(7) could in principle be solved without assuming
  separability, however, the resulting solution is much less
  transparent and we will therefore not pursue this path here.}

\begin{figure}
        \includegraphics[angle=0,width=0.49\textwidth]{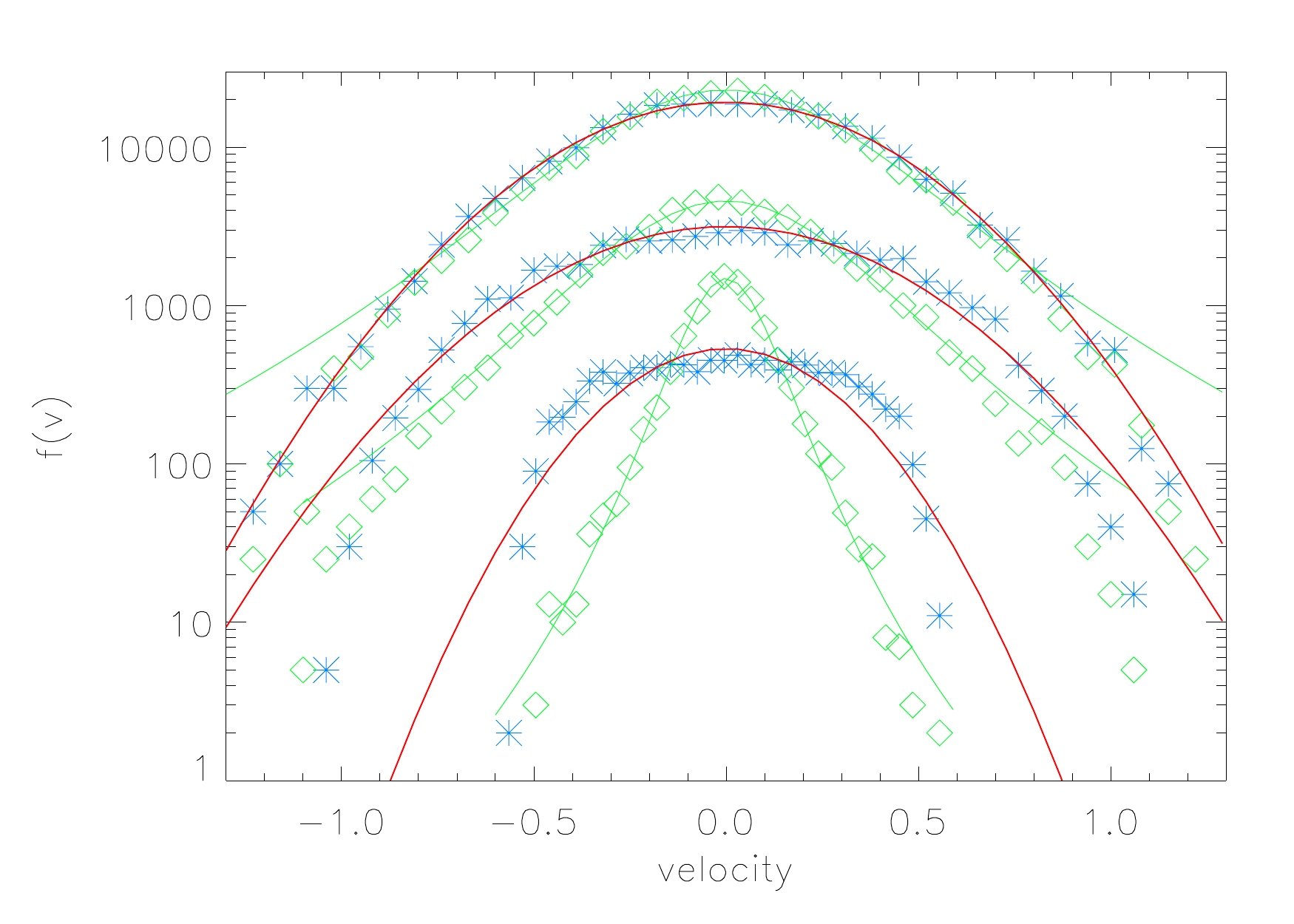}
\caption{The radial (blue stars) and tangential (green diamonds) velocity distribution at 3
  different radii (corresponding to $\gamma = -1.7, -2.4, -3.0$) after repeated
  {\em kick-flow} perturbations of the particle energies. The figures are
  shifted vertically to improve readability. The radial distributions
  are all fitted with the shape in Eq.~(\ref{eq:frad}) (thick red lines), and the
  tangential distributions are all fitted with the shape in
  Eq.~(\ref{eq:ftan}) (thin green lines). It is clear that the
low-velocity region is well fitted, whereas the high-velocity region
is not.}
\label{fig:fig5}
\end{figure}

\section{Conclusion}
We derive the equation involving only the radius and the radial
component of the velocity distribution function by integrating the
Boltzmann equation over both the tangential velocities and the spatial
angles. We solve this equation under the assumption that the
distribution function is separable in the radial and tangential
velocities. These solutions indicate that the radial distribution
function is close to a Gaussian shape, Eq.~(\ref{eq:frad}), whereas the tangential is
close to a Tsallis shape, Eq.~(\ref{eq:ftan}).

We compare the solution of the radial distribution function with the
results of numerical simulations, and we find that for velocities
smaller than approximately 0.7 times the escape velocity, the solution
fits rather well. In principle this is a prediction 
with zero free parameters, and hence the agreement with simulated
data is not entirely trivial.
Thus, for small velocities (which is the dominant
component) the full velocity distribution function can be derived
analytically.

\section*{Acknowledgement}
It is a pleasure to thank Martin Sparre for providing data for the
figures.  We thank the anonymous referee for suggestions which
improved the paper. We thanks Gary Mamon for constructive comments. 
This project is partially funded by the Danish
council for independent research, under the project ``Fundamentals of
Dark Matter Structures'', DFF – 6108-00470.

\label{lastpage}

\end{document}